\documentstyle[12pt]{article}
\begin{document}

\title{{\Large {\bf THERMODYNAMIC APPROACH TO THREE-SITE ANTIFERROMAGNETIC ISING
MODEL IN CHAOTIC REGION }}}
\author{N.S. Ananikian\thanks{
e-mail: ananik@jerewan1.yerphi.am} , S.K. Dallakian\thanks{
e-mail: saco@atlas.yerphi.am} \\
%EndAName
{\normalsize Department of Theoretical Physics, Yerevan Physics Institute,}\\
{\normalsize Alikhanian Br.2, 375036 Yerevan, Armenia}}
\maketitle

\begin{abstract}
The chaotic properties of the three-site antiferromagnetic Ising model on
Husimi tree are investigated in magnetic field. Macroscopic quantity of
three-site antiferromagnetic Ising model is generated by one dimensional
map. It is shown that in certain parameter setting strange attractors of
this map exhibit multifractal scaling. By applying thermodynamic formalism
we find nonanalyticity in free energy as well as in entropy. We show that
the temperature of phase transition depends on parameter of model and varies
from negative to positive value.
\end{abstract}

\newpage

\begin{center}
1. INTRODUCTION
\end{center}

One of the formalism to analyze nonlinear physics having complicated fractal
objects and strange attractors is the thermodynamic approach\cite{A4,A22}.
The advantage of this formalism is that the number of degrees of freedom is
usually enormously large and their properties are described by
thermodynamical function which contain most of the relevant information
about macroscopic systems. In order to characterize these macroscopic
systems order parameter has been sought. One of this order parameters is
Lyapunov exponents. The tool with which we can measure Lyapunov exponents
and other dynamical quantities is similar to the thermodynamic formalism in
equilibrium statistical mechanics. By using thermodynamic formalism many
studies have been done on strange attractors and it has been found
nonanalitisity in free energy function and scaling in the distribution of
Lyapunov exponents\cite{A21,DD}. Besides there are many other relationships
between Ising-like systems and thermodynamic formalism\cite{A19}.

In this paper we investigate the three-site antiferromagnetic Ising spin
model (TSAI) and show how to obtain connection between the TSAI model and
thermodynamic formalism. The reason for studying the Ising model with
multisite interaction is that it plays an important role for investigations
of real physical systems, such as binary alloys \cite{A11}, classical fluids 
\cite{A12}, solid $^3He$ \cite{A13}, lipid bilayers \cite{A14}, and rare
gases \cite{A15}. Recently the multisite interaction Ising model on Husimi
tree has been investigated. It is shown, that this approach yields
qualitatively good approximation for the ferromagnetic phase diagrams than
conventional mean-field calculation \cite{A5}. In contradiction to
ferromagnetic case, when we change the sign of three-site interaction the
situation changes drastically. In certain values of temperature and magnetic
field the TSAI model has nontrivial thermodynamic limit and as a consequence
the magnetization exhibits chaotic behavior\cite{A7,AA8}. In this case the
magnetization is not an order parameter. In this paper we introduce Lyapunov
exponent as the order parameter in chaotic region using thermodynamic
formalism.

With the ''thermodynamic formalism'' we investigate TSAI model in the
chaotic region and describe its chaotic properties via the invariants
characterizing fractal sets (e.g. strange attractors). In particular, we
obtain the entropy and free energy function and focus on whether the
thermodynamical quantities have phase transition \cite{A21}. It is, in
general, hard to determine such behavior unambiguously by numerical methods
if one does not have further arguments or exact solutions (as in the case of 
$x\to 4x(1-x)$). However, there is a possibility to remedies these
deficiencies if one considers the characteristic Lyapunov exponents as an
order parameter, which will differ in two phases. In this paper we
numerically calculate this order parameter as a function of the
''temperature'' and show where the phase transition occurs.

The contents of the remainder of this paper is as follows. TSAI\ system on
Husimi tree and recursion relation is presented in Sec. 2. In Sec 3. there
is a discussion of the property of TSAI model in thermodynamic limit. In
Sec. 4 we present the TSAI model in the case of ''fully developed chaos''
and obtain the exact connection between this model and chaotic attractors.
By using the ''thermodynamic formalism'', the phase transition in terms of
Lyapunov exponents, free energy and entropy function is analyzed. Finally,
in Sec. 6 we summarize our results and comment on their implications for the
study of other systems.

\begin{center}
2. TSAI MODEL ON\ HUSIMI\ TREE\ AND\ RECURSION RELATION\ 
\end{center}

The advantage of the Husimi tree is that for models formulated on it, an
exact recursion relations can be obtained. The pure Husimi tree \cite{A17},
shown in Fig.1, is characterized by $\gamma $, the number of triangles which
goes out from each site and $n$, the number of generation . The
0th-generation is a single site from which come out $\gamma $ triangles. All
subsequent generation comes out by gluing up $\gamma -1$ triangles to each
free sites of previous generation.

The TSAI model in a magnetic field is defined by the Hamiltonian one 
\begin{equation}
\label{R1}H=-J_3^{^{\prime }}\sum_{\triangle }{\sigma }_i{\sigma }_j{\sigma }%
_k-h^{^{\prime }}\sum_i{\sigma }_i, 
\end{equation}
where ${\sigma }_i$ takes values $\pm 1$, the first sum goes over all
triangular faces of the Husimi tree and the second over all sites.
Additionally, we use the notation $J_3=\beta J_3^{^{\prime }}$, $\ h=\beta
h^{^{\prime }}$, $\beta =1/kT$, where h is the external magnetic field, T is
the temperature of the system and $J_3<0$ corresponding to an
antiferromagnetic coupling (in all our numerical calculations we put $%
J_3^{^{\prime }}=-1$).

The partition function will be written as 
\begin{equation}
\label{R2}Z=\sum_{\{\sigma \}} exp\left\{ J_3\sum_{\triangle}{\sigma}_i{\
\sigma}_j {\sigma}_k+h\sum_i{\sigma}_i \right\} , 
\end{equation}
where the summation goes over all configurations of the system.

When the Husimi tree is cut apart at the base site, it separates into $%
\gamma $ identical branches. The partition function can be written as
follows: 
\begin{equation}
\label{R3}Z_n=\sum_{\{\sigma _0\}}exp\left\{ h\sigma _0\right\} {[g_n({%
\sigma _0})]}^{\gamma -1}, 
\end{equation}
where ${\sigma }_0$ are spins of base site, $n$ is the number of generations
($n\to \infty $ corresponds to the thermodynamic limit). Each branch, in
turn, can be cut along any site of the 1th-generation which is the nearest
to the central site. The expression for $g_n({\sigma }_0)$ can therefore be
rewritten in the form 
\begin{equation}
\label{R5}g_n({\sigma }_0)=\sum_{\{\sigma _1\}}exp\left\{ J_3\sum_{\triangle
}{\sigma }_0{\sigma }_1^{(1)}{\sigma }_1^{(2)}+h\sum_{j=1,2}{\sigma }%
_1^{(j)}\right\} {[g_{n-1}({\sigma }_1^{(1)})]}^{\gamma -1}{[g_{n-1}({\sigma 
}_1^{(2)})]}^{\gamma -1}. 
\end{equation}
From Eq.(\ref{R5}) we easily obtain 
$$
g_n(+)=e^{J_3+2h}g_{n-1}^{\gamma -1}(+)g_{n-1}^{\gamma
-1}(+)+2e^{-J_3}g_{n-1}^{\gamma -1}(+)g_{n-1}^{\gamma
-1}(-)+e^{J_3-2h}g_{n-1}^{\gamma -1}(-)g_{n-1}^{\gamma -1}(-), 
$$
$$
g_n(-)=e^{-J_3+2h}g_{n-1}^{\gamma -1}(+)g_{n-1}^{\gamma
-1}(+)+2e^{J_3}g_{n-1}^{\gamma -1}(+)g_{n-1}^{\gamma
-1}(-)+e^{-J_3-2h}g_{n-1}^{\gamma -1}(-)g_{n-1}^{\gamma -1}(-). 
$$
We introduce the following variable: 
\begin{equation}
\label{R6}x_n=\frac{g_n(+)}{g_n(-)}. 
\end{equation}
For $x_n$ we can then obtain the recursion relation 
\begin{equation}
\label{R7}x_n=f(x_{n-1}),\qquad f(x)=\frac{z{\mu }^2x^{2(\gamma -1)}+2\mu
x^{\gamma -1}+z}{{\mu }^2x^{2(\gamma -1)}+2z\mu x^{\gamma -1}+1}, 
\end{equation}
where $z=e^{2J_3},\quad \mu =e^{2h}\ and\ 0\le x_n\le 1.$ The function $f(x)$
is unimodal: it is continuous, continuously differentiable, and has one
maximum $x^{*}$ in $[0,1]$. Note that $f(x^{*})=1$ for any $\gamma $, $h$
and $T$. This function is nonhyperbolic (hyperbolicity for $1D$ maps means
that $1<|f^{\prime }|<\infty $ in all points) and maps the interval $[0,1]$
onto $[z,1]$.

Through $x_n$, obtained by Eq.(\ref{R7}), one can express the magnetization
of the central base site: 
\begin{equation}
\label{R8}m_n=\langle \sigma _0\rangle =\frac{e^hg_n^\gamma
(+)-e^{-h}g_n^\gamma (-)}{e^hg_n^\gamma (+)+e^{-h}g_n^\gamma (-)}=\frac{%
e^{2h}x_n^\gamma -1}{e^{2h}x_n^\gamma +1}, 
\end{equation}
and other thermodynamic parameters, so we can say that the $x_n$ determines
the states of the system. For example, at high values of temperatures (see
next subsection) the recursion Eq.(\ref{R7}) tends to a fixed point and
therefore the system has an appointed magnetization $m$. For the free energy
function (using Eqs.(\ref{R3}), (\ref{R5}), (\ref{R6}), (\ref{R7})), we
obtain 
\begin{equation}
\label{R9}F_n=-\frac \gamma 3J_3-\frac 12lnt+\frac{(\gamma -1)}%
2ln(zt^2+2t+z)(t^2+2zt+1)-\frac{(2\gamma -3)}3ln(zt^3+3t^2+3zt+1), 
\end{equation}
where $t=\mu x_n^{(\gamma -1)}$. In deriving Eq.(\ref{R9}) we have used the
following relation: 
\begin{equation}
\label{R10}F=-\frac{2\gamma -3}3F_{Cayley}, 
\end{equation}
where $F_{Cayley}$ is the free energy function of the triangular Cayley
tree. It is easily seen that the expression for magnetization given by Eq.( 
\ref{R8}), can be obtained by differentiating the free energy function of
Eq.(\ref{R9}) with respect to the magnetic field $h$. It is interesting to
mention that Eq.(\ref{R10}) is the generalization of the results obtained in
Ref.\cite{A18,AG}.

\begin{center}
3.THERMODYNAMIC\ LIMIT
\end{center}

Let us consider the magnetization of the central base site. In order to
achieve thermodynamic limit we tend the number of generation to infinity ($%
n\to \infty $). If we set $\gamma =4$ in Eqs.(\ref{R7}),(\ref{R8}) and vary
the temperature and magnetic field , then for sufficiently large $T$ we see
that function $f(x)$ has a stable fixed point at every value of $h$ and
therefore map $x_n=f(x_{n-1})$ attracts every point to $x^{*}=f(x^{*})$ in
the thermodynamic limit ($n\to \infty $). At such temperature the
magnetization $m$, is a well defined function of $h$ (see Fig.2a) in
thermodynamic limit. Then we lower $T$, at some values of $h$ the point $%
x^{*}=f(x^{*})$ becomes unstable , but now two new stabile points $x_1,x_2$
arise in function $f(f(x))$ , which is an attracting periodic orbit of
period $2$ for map $x_n=f(x_{n-1})$ and we find that there is a single
bubble in the plot of $m$ versus $h$ (Fig.2b). As we continue to lower $T$,
new bubbles are formed as parts of the old bubbles (Fig.2c, attracting
periodic orbits of period $2^N$), and for still lower $T$ s we reach a
region where for intermediate values of $h$ we have chaos, period-three
windows, etc. (see Fig.2d).

One can say that the reason for this is that we have the presence of
frustration effects and taking into account antiferromagnetic nature of the
three-site interaction identifies two different magnetization in fixed field
and temperature with magnetization of sublattice, as in antiferromagnetic
phase and chaotic region with spin glass phase\cite{A10}. But as one can
note from our calculation different magnetization corresponds to different
sample with different number of generation. Therefore the appearance of a
bubble dues to fact that our system does not have trivial thermodynamic
limit and in that limit different samples with different number of
generation are not macroscopically equivalent systems.

Obviously magnetization can't be an order parameter in chaotic region. How
we can identify the appearance of chaotic region? From the theory of
dynamical system we know that Lyapunov exponents are good order parameter
and become positive when attractors of the system become chaotic or strange%
\cite{A19}. A statistical-thermodynamic method for the description of
fluctuation of the Lyapunov exponents has been introduced, using a partition
function\cite{A4,A22}. Later on we use thermodynamic formalism in order to
characterize TSAI\ system in chaotic region.

It is interesting to note that similar chaotic behavior has been found in
other frustrated hierarchical lattices\cite{A10,S11}. The Husimi tree like
other hierarchical lattices are effectively infinite dimensional. But in
Husimi or Bathe like lattices the number of neighborhood remain constant in
contradiction to other hierarchical lattices for which in thermodynamic
limit surface sites interacts with infinitely many neighborhood. Another
difference is in nonequivalentce of the lattice sites. In Husimi or Bethe
like lattices the sites lying deep inside lattices are equivalent\cite{S12}.

\begin{center}
4.TSAI SYSTEM AND ''THERMODYNAMIC FORMALISM''
\end{center}

How to apply the ''thermodynamic formalism'' to the TSAI system? For this we
need a natural partition and that is provided by the cylinders (we follow
here Ref. \cite{A21}).

The recursion function Eq.(\ref{R7}) with $\gamma =4$ has the following form 
\begin{equation}
\label{R26}f(x)=\frac{z{\mu }^2x^6+2\mu x^3+z}{{\mu }^2x^6+2z\mu x^3+1}. 
\end{equation}
Simultaneously considering the following system of equation, 
\begin{equation}
\label{R12}\cases{f(x_0)=x_0\cr f(1)=x_0\cr}, 
\end{equation}

we obtain 
\begin{equation}
\label{R31}z =\frac{{\mu}^{-2/3}+{\mu}^{-8/3}-2{\mu}^{-1}} {1+{\mu}^{-2}-2{%
\mu}^{-5/3}}, \quad x_0={\mu}^{-2/3}. 
\end{equation}

For a crisis map (Eqs.(\ref{R26}), (\ref{R31})) we want to describe the
scaling properties of the attracting set which in this case is the entire
interval $I$: $[x_0,1]$ (Fig.3). For an index $n$, $I$ is partitioned into $%
2^n$ intervals or n-cylinders, these being the segments with identical
symbolic-dynamics sequences of length $n$ taken with respect to the maximum
point $x^{*}=c$. The inverse function of Eq.(\ref{R26}), $h=f^{-1}$, has two
branches, $h_{-1}$ and $h_1$ as shown in (Fig.3) and the $n$-cylinders are
all the nth-order preimages of $I$. The length of the cylinders is denoted
by $l_{\epsilon _1,\epsilon _2,\dots \epsilon _n}\equiv h_{\epsilon _1}\circ
h_{\epsilon _2}\circ \dots \circ h_{\epsilon _n}(I)$ where $\epsilon \in
\{-1,1\}$.

The partition function $Z(\beta )$ is defined \cite{A21} as 
\begin{equation}
\label{R35}Z_n(\beta )=\sum_{\epsilon _1,\epsilon _2,\dots \epsilon
_n}l_{\epsilon _1,\epsilon _2,\dots \epsilon _n}^\beta =\sum_{\epsilon
_1,\epsilon _2,\dots \epsilon _n}e^{-\beta ln{l_{\epsilon _1,\epsilon
_2,\dots \epsilon _n}}}, 
\end{equation}
where $\beta \in (-\infty ,\infty )$ is a free parameter, the inverse
''temperature''. In the limit $n\to \infty $ the sum behaves as 
\begin{equation}
\label{R36}Z_n(\beta )=e^{-n\beta F({\beta })}, 
\end{equation}
which defines the free energy, $F(\beta )$. The entropy $S(\lambda )$ is the
Legendre transform 
\begin{equation}
\label{R37}S(\lambda )=-\beta F(\beta )+{\lambda }{\beta }, 
\end{equation}
where the relation between $\lambda $ and $\beta $ is found from 
\begin{equation}
\label{R38}\lambda =\frac d{d\beta }{(\beta F(\beta ))},\quad \beta (\lambda
)=S^{\prime }(\lambda ), 
\end{equation}
and these have the following meaning: In the limit $n\to \infty $, $%
e^{nS(\lambda )}$ is the number of cylinders with length $l=e^{-n\lambda }$
or, equivalently, with Lyapunov exponent $\lambda $, which we consider as an
order parameter for TSAI model . The Hausdorf dimension of the set of points
in $I$ having Lyapunov exponent $\lambda $ is $S(\lambda )/\lambda $ \cite
{A21}.

By using the Eqs.(\ref{R26}), (\ref{R31}), (\ref{R35}), (\ref{R36}) we can
numerically calculate the free energy at different value $\mu $ which is
shown in Fig.4. One can see from Fig.4a that the free energy have a
singularity around $\beta _c=-1$ at low values of $\mu $ while at high
values of $\mu $ the free energy have a singularity around $\beta _c=1$ (Fig
4b), which shows the existence of the phase transition of first order in
this regions of $\beta $.

How can the transition be determined accurately?

Let us consider the characteristic Lyapunov exponent as an order parameter
which will differ in the two phases. Fig.5 shows this order parameter for
different $\mu $ and sizes of the system, corresponding to $n=7,9,11,13$.
The curves converge towards a line and the result is the first order
transition, whereas in a class of maps close to $x\to 4x(1-x)$ a phase
transitions of the first order occur only with negative $\beta _c$. We would
like to mention that it is hard to determine the critical ''temperature'' of
phase transition by numerical methods with high precision. Consequently, the
obtained value of critical ''temperature'' is approximate. In connection
with this it is important to note, that there is a literature on phase
transition in fully developed chaotic maps with neutral points leading to
intermittent dynamics \cite{DD}, that claims the existence of a phase
transition in a free energy at ''temperature'' $=1$.

Large deviations of fluctuations of a Lyapunov exponents can be described by
the entropy. To consider the above results in terms of the entropy function $%
S(\lambda )$, let us first discuss the general appearance of the entropy
function. First of all, it should be positive on some interval $[\lambda
_{min},\lambda _{max}]$. The value $\lambda =ln2$ must belong to that
interval, which follows from the fact that the sum of the lengths of all
cylinders on a given level is $1$. Secondly it is often found that the
values of $\lambda _{min}$ and $\lambda _{max}$ are given by the logarithms
of the slopes at the origin .

The precise form of the entropy function is, as mentioned in the
introduction, not easy to obtain with great accuracy. The existence of a
first order phase transition implies that there should be a straight line
segment in $S(\lambda )$ and the slope of the line equal to $\beta _c$. This
scenario is seen in Fig. 6. The curve in the figure corresponds to $n=13$ Of
course, with the finite-size data, it is impossible to determine the
straight line segment in $S(\lambda )$ and with increasing $n$, the straight
line will be increased.

\begin{center}
6. CONCLUSION
\end{center}

In this paper we have investigated the TSAI model by approximating it with a
Husimi tree structures in an external magnetic field, and a strong
connection with results from the theory of dynamical systems including chaos
has been pointed out.

An exact connection between that statistical system and fully developed
chaotic attractors is obtained. Is is shown that in the chaotic region the
magnetization, which are the order parameter for ferromagnetic phase, must
be replaced by Lyapunov exponent in order to characterize TSAI model in
chaotic region. The chaotic properties of the antiferromagnetic multisite
system is described via the invariants characterizing a fractal set (e.g. a
strange attractor). It is shown that this system in the chaotic region
displays a phase transition. It is, in general, hard to determine such
behavior unambiguously by numerical methods if one does not have further
arguments or exact solutions. In this paper we have considered the
characteristic Lyapunov exponents as an order parameter, which will differ
in two phases. We've managed to calculate this order parameter as a function
of the ''temperature'' and approximately showed where the phase transition
will occur, since it describes transitions in the distribution of the
characteristic Lyapunov exponents. This phase transition in terms of free
energy and entropy function is also analyzed.

It is interesting to note that for $\gamma =3$ the above situation changes
dramatically and it will be very difficult to study numerically the chaotic
region of TSAI system in terms of the ''thermodynamic formalism''\cite{A7}.

On other hand, the study of chaotic statistical physical system has opened
new challenges for theories of stochastic processes, especially in the
direction of stochasticity of the vacuum in QCD \cite{A25,A8}. In this
direction interesting results for the $Z(Q)$ gauge model with a double
plaquette representation of action on the flat and generalized Bethe
lattices were obtained\cite{AA13,SS}. Note that it is possible to get a
chaotic region in the $Z(Q)$ gauge model with a three plaquette
representation of the action. The detailed investigations of these questions
will be published elsewhere.

\newpage

\begin{center}
Figure Captions
\end{center}

\vspace*{18cm}

Fig.1. Husimi tree with $\gamma =4$.

\newpage\ 

\vspace*{18cm}

Fig.2. Plots of $m$ versus $h^{\prime }$ for different temperatures $T$ ($%
\gamma =4$). a - $T=3$, b - $T=1.3$, c - $T=1.15$, d - $T=0.7$.

\newpage\ 

\vspace*{18cm}

Fig.3. The function of Eq.(\ref{R26}) for values of $\mu =105$ and $z$ given
by Eq.(\ref{R31}).

\newpage\ 

\vspace*{18cm}

Fig.4. The free energy $F(\beta )$, for different values of $\mu $. a - $\mu
=5$, b - $\mu =105$.

\newpage\ 

\vspace*{18cm}

Fig.5. The order parameter $\lambda (\beta )$ calculated for different sizes
of the system, corresponding to $n$ = 9, 11, 13 and different $\mu $. a - $%
\mu =5$, b - $\mu =105$.

\newpage\ 

\vspace*{18cm}

Fig.6. The entropy function corresponding to $n$ = 13 for different $\mu $.
a- $\mu =5$, b - $\mu =105$ .

\end{document}